\documentclass{iaus}
\usepackage{graphicx}
\usepackage{graphics}
\usepackage{lscape} 

\newcommand{\mnras}{MNRAS}
\newcommand{\aap}{A\&A}
\newcommand{\apj}{ApJ}   
\newcommand{\apjl}{ApJL} 
\def\simlt{\mathrel{\hbox{\rlap{\hbox{\lower4pt\hbox{$\sim$}}}\hbox{$<$}}}}
\def\simgt{\mathrel{\hbox{\rlap{\hbox{\lower4pt\hbox{$\sim$}}}\hbox{$>$}}}}

\title[Magnetars and GRBs] 
{Magnetars and Gamma Ray Bursts}

\author[Niccol\`o Bucciantini]   
{Niccol\`o Bucciantini}

\affiliation{INAF, Osservatorio Astrofisico di Arcetri, L.go Fermi 5, 50125, Firenze, Italia\\ email: {\tt niccolo@arcetri.astro.it} \\[\affilskip]}

\pubyear{2012}
\volume{279}  
\pagerange{1--8}
\jname{Death of Massive Stars: Supernovae \& Gamma-Ray Bursts}
\editors{P.Roming, N. Kawai \& E. Pian, eds.}
\begin{document}

\maketitle

\begin{abstract}
In the last few years, evidences for a long-lived and sustained
engine in Gamma Ray Bursts (GRBs) have increased the attention to the
so called {\it millisecond-magnetar} model, as a competitive
alternative to the standard collapsar scenario. I will review here the key aspects of the {\it millisecond
magnetar} model for Long Duration Gamma Ray Bursts (LGRBs). I will
briefly describe what  constraints, present observations put on any
engine model, both in term of energetic, outflow properties, and the
relation with the associated Supernova (SN). For each of these I will
show how the millisecond magnetar model satisfies the requirements, what
are the limits of the model, how can it be further tested, and what
observations might be used to discriminate against it. I will also
discuss numerical results that show the importance of the
confinement by the progenitor star in explaining the formation of a
collimated outflow, how a detailed model for the evolution of the
central engine can be built, and show that a wide variety of explosive
events can be explained by different magnetar parameters. I will
conclude with a suggestion that magnetars might be at the origin of
the Extended Emission (EE) observed in a significant fraction of Short GRBs.

\end{abstract}

\firstsection 
\section{Introduction}

The key idea behind the so called  {\it magnetar} model  for LGRBs (\cite{usov92},\cite{thom94},\cite{wheeler00},\cite{thom04}) 
assumes that the collapse of the core of a massive star at the
end of its life leads to a
rapidly rotating proto-neutron star (NS) (period $\sim 1$ ms), with a strong
surface magnetic field (B $\ge 10^{15}$ G). There are reasons to
belive that this two properties might be related by dynamo
processes (\cite{tho95,tho96}). The maximum energy that can be stored in
a rotating NS is $\sim 2\times 10^{52} {\rm erg}$, and the typical
timescale over which this energy can be extracted and delivered to the
surrounding medium is $\sim 100{\rm s}$ for a magnetic field $\sim
3\times 10^{15}{\rm G}$. These energies and timescales are compatible
with almost every LGRB observed. Moreover, the formation of a
proto-neutron star, is fundamental for a successful supernova
explosion, NSs are known to produce relativistic outflows (\cite{bucciantini08}), and there
are evidences that massive stars might not necessarily end their life
forming Black Holes (\cite{mun06,bel08,gae05,wht08,gae06,vin08,mor07}). Indeed the rotational energy available in a
millisecond proto-NS, is more than sufficient to unbind the
envelope of even a 40-60 $M_\odot$ star (\cite{metzger+11}), and, for a strong
magnetic field, mass losses can outpace accretion (\cite{dessart+08}), and the wind might
halt the fallback of marginally bound ejecta (\cite{b09})

\section{Energetics and timescales for proto-NS winds}

Once a proto-NS is formed, it will cool
via neutrino emission in a typical Kelvin-Helmholtz timescale
$t_{KH}\sim 10-100$ s (\cite{pons99}). Neutrinos  deposit heat in a
neutrinosphere and, at about 500-800 ms after core bounce, the density
surrounding the proto-NS can drop to the point where a neutrino driven
wind develops (\cite{arc07,thom01}). For proto-NS with pulsar like magnetic fields, this wind
carries negligible energy (\cite{thom01}). However, for magnetar like magnetic fields,
the wind is magnetocentrifugally accelerated and far more
energetic (\cite{thom04}). As the neutrino luminosity decreases, and the mass loss
rate drops, the wind becomes progressively more magnetized eventually
reaching relativistic speeds (\cite{thom04,b05}).

In a recent paper \cite[Metzger et al. (2011)]{metzger+11} have developed a full model
for the spin-down evolution of a proto-NS, taking into account the
neutrino cooling, the magnetic torque by the wind, and the possible
inclination of the magnetic axis with respect to the spin axis. The
model can be used to estimate the energy flux in the wind and its
magnetization parameter $\sigma$ which is defined as the ratio of
Poynting flux over kinetic energy flux, and can be thought of as a proxy
for the maximum achievable Lorentz factor.   

Five phases can be distinguished in the overall evolution (\cite{metzger+11}):
\begin{itemize}
\item An early non relativistic phase, that lasts $\sim 1$ s after bounce, when
  the proto-NS is still hot and contracting, the wind is magnetized,
  but with typical terminal speeds $\sim 0.1c$.
\item An intermediate mildly relativistic phase  that lasts $\sim 10$ s
  after bounce. The proto-NS relaxes to a radius $\sim 20$ km, and
  begins to spin-down, the
  wind mass loss rate drops, and $\sigma $ increases from values $\sim
  1$ to $\sim 10$. The wind is now relativistic,  but still confined
  inside the progenitor.
\item The GRB phase, after the wind breaks out of the progenitor, and
  starts to accelerate in the circumstellar medium. Due to the
  decreased neutrino luminosity, and related mass loss rate,  the
  magnetization rapidly increases to values $\sigma \sim 10^2-10^3$, while
  energy losses are still high $\sim 10^{49}$ erg s$^{-1}$.  This
  phase lasts for about 100 s, when the neutrino luminosity drops
  sharply below the threshold to drive a barion loaded wind.
\item A late activity phase, that begins once the barion loaded wind
  ceases, to be replaced by a leptonic wind once the density in the
  magnetosphere drops below the threshold for pair production. The wind
  luminosity is smaller  $\sim 10^{48}$ erg s$^{-1}$, but the typical
  spin-down time is now longer $\sim 100 $ s, as expected to explain
  some aspects of the late activity observed in some LGRBs.
\end{itemize}

We want to emphasize here that energy losses are due to a magnetized
stellar wind, analogous to the Solar wind and other stellar winds, and
are not due to magnetic dipole radiation. Interestingly for $\sigma
\gg 1$ the two have the same value, but in the early and mildly
relativistic phases the energy losses in the wind can exceed even by a
factor 10 what can be estimated based on dipole radiation alone.

 \section{Collimation and acceleration of the outflow}

As opposed to accretion disks around black holes that are known to
power collimated outflow in the form of relativistic jets, magnetars
are supposed to produce essentially spherical outflows. Relativistic
outflows cannot self collimate (\cite{le01}), and at large distances from
the Light Cylinder, the wind structure should closely follow the split
monopole solution (\cite{b05}), where most of the energy flux is in the equatorial
region. Moreover in monopolar relativistic outflows, the terminal
Lorentz factor can only be as high as $\sigma^{1/3}$ (\cite{aro01}). This
is called the $\sigma$ problem. It is evident that the GRB outflow
that is observed, cannot coincide with the the steady state spherical
wind emerging directly from the proto-NS magnetosphere.

We know that, as the outflow emerges from the interior of the host
star, it collimates into a jet that
punches through the stellar envelope, creating a channel where
material coming from the central engine can flow (e.g.,
\cite{mat03}). Afterglow observations ({\it jet breaks};
\cite{rho99}) and GRB energetics (comparison of the total energy
derived from late radio afterglow observations with respect to the
prompt emission) confirm that a collimated flow is present.

Simple energy considerations demonstrate that the surrounding
stellar envelope provides an efficient confining medium even for a
very energetic proto-magnetar wind.  It is indeed the interaction with
the progenitor star that provides the collimating agent to channel the
spherical magnetar wind into a polar jet. By analogy to bipolar wind bubbles
(\cite{kg02},\cite{beg92}) the interaction of the wind from the
spinning-down magnetar with the surrounding star could facilitate
collimation. 
It has been shown, under different assumptions,  that once the interaction
with the surrounding progenitor is taken into account, this
can in fact occur (\cite{b07,b08,k08,b09}).

The physical picture is based on an analogy to the case of pulsar wind nebulae (PWNe)
(\cite{kom04,ldz04}): the magnetar wind is confined by the surrounding
(exploding) stellar envelope, as a result  a {\it magnetar wind
  nebula} (MWN) forms, where the wind magnetic field is compressed. If the toroidal magnetic field in the bubble is sufficiently
strong, due to the
{\it tube of toothpaste} effect
the bubble expands primarily in the polar direction while a negligible
amount of energy is transferred to the SN
envelope.

The issue of collimation is strictly related to the problem of the
acceleration because deviations from the strict monopole geometry can substantially enhance the
terminal Lorentz factor, as well as time dependent effects. In the millisecond magnetar, the collimation
of the outflow, and the formation of a wind nebula becomes a key
features if one wants to build a GRB engine. 
A key assumption here is that magnetic energy can be
efficiently dissipated/converted into kinetic energy
(\cite{aro01,kir01,kir03}).  This can happen in various ways. For an
oblique rotator, the striped magnetic field might reconnect and dissipate at the wind termination
shock of the nebula as it slows down (\cite{lyu05}). It is not unlikely
that instabilities and dissipation might be at work inside the nebula
itself or in the jet (\cite{beg92,spr08}). Modulation of the outflow by
the confining walls of the channel formed inside the progenitor migh
enhance magnetic to kinetic energy conversion (\cite{granot+11}).

Results show that the opening angle of the jet, as it punches trough
the star and later emerges into the circumstellar medium, is of order
of $5-10^\circ$, and appears to be independent of the dissipative
properties of the magnetar wind, as long as the magnetization inside
the MWN reaches equipartition. It is also shown that the outflow can
accelerate rapidly as soon as it emerges from the progenitor star.

\section{SN association and late activity}

It is now well established that  long-duration GRBs are associated 
with core-collapse SNe, in particular with the subclass of SNe Ic-BL (BL = broad
line) (\cite{woo06,del06,zha07}).  Interestingly, the converse is not true
(\cite{soderberg06,woo06}), and the search for orphan afterglows  shows
that within a high confidence level the hypothesis that every broad-lined SN harbors
a GRB  can be ruled out.  Moreover, even if the SNe associated with
long-duration GRBs, tend to be more luminous than the average sample,
they  are not
particularly unusual among the class of BL SN in terms of their
energies, photospheric velocities, and Ni masses.

It can be debated if this class of core-collapse SNe that are unusually energetic and
asymmetric (as revealed by spectra-polarimetry), and that produce
significant amounts of Ni, are powered by a diverse central engine (a
failed GRB), or if the GRB engine is only a possible outcome of the
conditions leading to those supernovae: are hypernovae due to a
GRB-like engine, or vice-versa? Higher energies, axisymmetry and Nikel
production are the three aspects that must be considered in the GRB-SN association. 

In the magnetar model nearly all of the spindown
energy of the neutron star escapes in the polar channel (\cite{k08,b09}).
  There is very little coupling between
the exploding star and the GRB engine.  This seems to apply
both to the low (dissipative) and high (non dissipative) $\sigma$
limits. The interesting
implication of this result is that a proto-magnetar powering a GRB
is unlikely to contribute significantly to energizing the SN shock as
a whole (although it clearly does so in the polar region), at least on
timescales $\simgt 1$ sec after core bounce. This is a key property of
magnetized outflows, and in principle it is not specific to a particular central
engine. Specifically, we suspect
that the same results will apply also to winds from accretion disks
(\cite{proga03}), that will likely escape via
the polar channel rather than transferring energy to the SN shock as
has been previously hypothesized (\cite{aro03}). 

Given the relatively {\it
  on-axis} viewing angle of observed GRBs, high velocity ejecta might be
observed; high velocity O and Ne can also be produced by the jet
blowing out stellar material that had been processed during stellar
evolution (\cite[Mazzali et al. 2006]{maz06}). A jet  might lead to unique observable signatures in the ejecta
at late times (as may be the case for Cas A; \cite[Wheeler et al. 2008]{whe08}). 

A separate issue relates  {\it magnetar} engines and the production of
excess $^{56}$Ni, that is observed in hypernovae. It has been shown that
the temperature at which explosive
nucleosynthesis of $^{56}$Ni happens ($\simgt 5 \,
10^9$ K; \cite{woo02}), is not attained even at relatively early
times. This happens because, by the time the jet-plume emerges outside the SN
shock, the density of the progenitor, into which it propagates, is $\sim 10^{4-5}$ g cm$^{-3}$. At
these densities Ni production requires a shock moving at nearly the
speed of light, significantly faster than what can be achieved at
these early times (\cite{k08,b09,met07}).  However $\sim
10^{-2} M_\odot$ of high speed ($v \simeq 0.1-0.2 \, c$) Ne and O can
be created, because these have lower threshold temperatures for
successful explosive nucleosynthesis.

As a side point, the specific angular
momentum required for a millisecond magnetar engine is $J\simeq
3\times 10^{15}R^2_{10}P_1^{-1}$ cm$^2$s$^{-1}$ (\cite{thom04}), is about a factor five
smaller that what is required for the formation of an accretion disk
for the black-hole accretion-disk model (\cite{mcf99}). This implies that if the
core has enough angular momentum to power a {\it collapsar} engine,
 then it has enough to create a millisecond magnetar.

The millisecond magnetar model for LGRBs is particularly interesting
in view of the so called {\it late activity}. Late activity manifests itself in the afterglow up to $10^{4-5}$
seconds after the prompt emission
(\cite{can05,vau05,cus05,nou06,obr06,wil06}), either as a shallow decay
or plateau
of the light curve, or with the presence of flares
(\cite{bur05,fal06,chi07}). Late activity requires a persistent engine
at times much longer than the typical duration of the prompt emission,
to provide continuous injection of energy.

In the millisecond magnetar model late time injection can take the
form of a
leptonic wind, not dissimilar to standard pulsars. The amount of energy that can be released in this scenario,
even if smaller than the prompt emission, can produce the shallow decay
phase that is observed (\cite{yu07,metzger+11}).

More intriguing is the presence of flares, that can carry a
substantial fraction of energy compared to the prompt emission.
 A possibility is that these flares are the signature
of magnetic readjustments, within the proto-magnetar, that give rise to
busting activity not dissimilar to what is observed during giant
bursts in SGRs
(\cite{tho95,tho96,woo04,mer08}). The magnetic energy stored in
canonical magnetars is smaller than the rotational energy required for
a GRB engine, however
the internal magnetic field might be much higher than the surface
value (\cite{bra06,bra08}).

\section{Validating the model for LGRBs}

As shown before the magnetar model can reproduce many aspects of the
observed phenomenology in LGRBs. It is interesting to evaluate if and
how one can distinguish a magnetar from a different engine (i.e. a
Black Hole). Unfortunately, the dynamical properties of magnetized
outflows, once the value of Lorentz factor and of $\sigma$ are set,
are largely independent on the conditions at injection. A more
promising discriminant might be the composition: in particular within
the magnetar model one expects a transition from a barion loaded wind
to a leptonic dominated outflow at $\sim 100 $ sec after bounce.

Perhaps the bigger discriminant is the available energy. A magnetar
can store at most a few times $10^{52}$ erg of energy. The detection
of a GRB with higher total energy, could rule out a magnetar as its
engine. Determining, with accuracy, the total energy of a GRB is non
trivial. The prompt emission, must be corrected for beaming (\cite{cenko+10}), and
off-axis effects (\cite{vaneerten+10}), while the late radio emission is often
assumed to originate from a Sedov phase to be converted into a kinetic
energy (\cite{berger+10}). There is a small set of very energetic GRBs
(\cite{cenko+10}) that are marginally compatible with a magnetar
engine. However for the vast majority, and for those for which we have
good data, the inferred energies are always a few $10^{51}$ ergs (\cite{berger+10}).

There is also a set of GRBs with a long prompt emission, characterized
by several events lasting $\sim 100 $ sec and separated by quiescence
periods of about $200-400$ sec. Unlike for a BH scenario where one
might invoke bursty mass accretion, the magnetar spin-down is smooth.
However the gamma-ray luminosity might not be a good tracer of the
energy injection, depending on the efficiency of particle acceleration
in the outflow. In the recent paper by \cite[Metzger et al. (2011)]{metzger+11}) it was shown that
several expected correlations, like the Amati relation can
be recovered in the magnetar model assuming magnetic dissipation to be
at the origin of the radiation mechanism.

\section{Short GRBs with extended emission}

The standard LGRB/SGRB dichotomy has recently been challenged by
several `hybrid' events that conform to neither class
(e.g.~\cite{zha07}; \cite{Bloom+08}). All together
approximately $\sim 1/4$ of {\it Swift} SGRBs are accompanied by extended
X-ray emission lasting for $\sim 10-100$ s with a fluence $\gtrsim$
that of the GRB itself (see \cite{norris06} and \cite{perley09}
for a compilation of events).   The hybrid nature and common
properties of these events (`Short GRB' + $\sim$ 100 s X-ray tail)
have motivated the introduction of a new subclass: Short GRBs with
Extended Emission (SGRBEEs).  It was moreover recently discovered that
some SGRBs are followed by an X-ray `plateau' ending in a very sharp
break (GRB 980515; \cite{rowlinson10}; \cite{Troja+08};
\cite{Lyons+10})  difficult to explain by circumstellar
interaction alone.  Although the connection of this event to SGRBEEs
is unclear, it nevertheless provides additional evidence that the
central engine is active at late times. The long duration and high
fluence of the extended emission of SGRBEEs poses a serious challenge
to the NS merger scenario, because in this model both the prompt and
extended emission are necessarily powered by black hole accretion.  It
is in particular difficult to understand how such a high accretion
rate is maintained at very late times. \cite{met08} recently
proposed that SGRBEEs result from the birth of a rapidly spinning
proto-magnetar, created by a NS-NS merger or the AIC of a WD.  In this
model the short GRB is powered by the accretion of the initial torus
(similar to standard NS merger models), but the EE is powered by a
relativistic wind from the proto-magnetar at later times, after the
disk is disrupted.  Although a NS remnant is guaranteed in the case of
AIC, the merger of a double NS binary could also leave a stable NS
remnant. The interaction of the relativistic proto-magnetar wind with
the expanding ejecta was investigated by \cite[Bucciantini et al. (2012)]{b12}, with a focus on the confining role of the ejecta and its dependence on the wind power, and on the ejecta mass and density profile. 
The model thus predicts a class of events for which the EE is observable with no associated short GRB.  These may appear as long-duration GRBs or X-Ray Flashes unaccompanied by a bright supernova and not solely associated with massive star formation, which may be detected by future all-sky X-ray survey missions.

\begin{discussion}

\discuss{B. Zhang}{So you don't believe the magnetar model for
  Superluminous SNe}

\discuss{N. Bucciantini}{A millisecond magnetar might have enough
  energy to power a  Superluminous SN, but this energy, extrtacted by
  a magnetized wind, is only weakly coupled with the SN shock. I do
  not think that a magnetar can energize a SN shock, early enough to
  drive the nucleosynthesys of $\sim 0.5 M_\odot$ of $^{56}$Ni. }

\discuss{B. Zhang}{For NS-NS making a magnetar have you considered how
a supermassive magnetar form? What kind of NS equation of state is needed. }

\discuss{N. Bucciantini}{For NS-NS merger resulting in a long lived
  magnetar, one needs peculiar conditions: two low-mass NSs must be
  involved; a few tenths $M_\odot$ must be lost either by strong
  neutrino driven winds or during the merger itself; the EoS must be
  particularly stiff. The existence of a 2$M_\odot$ NS suggests that
  the EoS might allow for massive magnetars.}

\discuss{S. Moiseenko}{In our simulations of magneto-rotational
  supernovae explosion we found that magnetic field can reach the
  values $10^{14}-10^{15}$ G, but it is chaotic magnetic field which
  can be reduced in a short time due to reconnection. How to make a
  neutron star with so strong magnetic field?}

\discuss{N. Bucciantini}{It is true that MRI, and other instabilities
  migh enhance strongly the magnetic field, but this happens at small
  scales, and the resulting field is mostly chaotic and tends to
  dissipate rapidly. One of the  ideas behind the origin of the strong magnetar
  magnetic field, is that dynamo processes are at work. The key idea
  here is that a mean field dynamo operates. Investigating the
  possibility of mean field dynamo requires a full 3D geometry, with
  enough resolution to properly sample the parameter space in term of
  viscosity and resistivity. 2D simulation will all be subject to
  Cowling antidynamo theorem, so they can never lead to large scale
  fields. To my knowledge investigation of MHD
  Supernovae in the full 3D regime is very demanding and quite
  limited. We do observe magnetar, so nature must find a way to
  produce them.}

\end{discussion}

\end{document}